\documentclass{JINST}

\newcommand{\degree}{^{\circ}}

\title{Measurement of the effect of non ionising energy losses on the leakage current of silicon drift detector prototypes for the LOFT satellite}

\author{E.~Del Monte$^{a,b}$\thanks{Corresponding
author.}, A.~Rachevski$^c$, G.~Zampa$^c$, N.~Zampa$^c$, P.~Azzarello$^d$, E.~Bozzo$^d$, R.~Campana$^{e,f}$, S.~Diebold$^g$, Y.~Evangelista$^{a,b}$, E.~Perinati$^g$, M.~Feroci$^{a,b}$, M.~Pohl$^h$, A.~Vacchi$^c$\\
\llap{$^a$}INAF - Istituto di Astrofisica e Planetologia Spaziali, \\
  Via Fosso del Cavaliere 100, I-00133 Roma, Italy\\
\llap{$^b$}INFN - Istituto Nazionale di Fisica Nucleare, Sezione di Roma Tor Vergata,\\
  Via della Ricerca Scientifica 1, I-00133  Roma, Italy\\
\llap{$^c$}INFN - Istituto Nazionale di Fisica Nucleare, Sezione di Trieste,\\
  Padriciano 99, I-34149 Trieste, Italy\\
\llap{$^d$}ISDC - Data Centre for Astrophysics, Universit\'{e} de Gen\`{e}ve,\\
  Chemin d'Ecogia 16, CH-1290 Versoix, Switzerland\\
\llap{$^e$}INAF - Istituto di Astrofisica Spaziale e Fisica Cosmica, Bologna,\\
  Via Piero Gobetti 101, I-40129 Bologna, Italy\\
\llap{$^f$}INFN - Istituto Nazionale di Fisica Nucleare, sezione di Bologna,\\
  Viale B. Pichat, 6/2, I-40127 Bologna, Italy\\
\llap{$^g$}IAAT - Institut f\"{u}r Astronomie und Astrophysik T\"{u}bingen, Universit\"{a}t T\"{u}bingen,\\
  Sand 1, 72076 T\"{u}bingen, Germany\\
\llap{$^h$}DPNC - D\'{e}partement de Physique Nucl\'{e}aire et Corpusculaire, Universit\'{e} de Gen\`{e}ve, \\
  Quai Ernest-Ansermet 24, CH-1211 Gen\`{e}ve, Switzerland\\
  E-mail: \email{ettore.delmonte@iaps.inaf.it}}

\abstract{The silicon drift detectors are at the basis of the instrumentation aboard the {Large Observatory For x-ray Timing} (LOFT) satellite mission, which underwent a three year assessment phase within the ``Cosmic Vision 2015 -- 2025'' long-term science plan of the European Space Agency. Silicon detectors are especially sensitive to the displacement damage, produced by the non ionising energy losses of charged and neutral particles, leading to an increase of the device leakage current and thus worsening the spectral resolution.

During the LOFT assessment phase, we irradiated two silicon drift detectors with a proton beam at the Proton Irradiation Facility in the accelerator of the Paul Scherrer Institute and we measured the increase in leakage current. In this paper we report the results of the irradiation and we discuss the impact of the radiation damage on the LOFT scientific performance.}

\keywords{X-ray detectors and telescopes; Radiation damage to detector materials (solid state)}

\begin{document}

\section{Introduction}
\label{sec:introduction}

Silicon Drift Detectors (SDDs) are solid-state devices proposed in the mid-1980s as high resolution position-sensitive detectors for fast ionizing particles and for spectroscopy of X-rays \cite{Gatti_Rehak_1984,Gatti_Rehak_2005}. Currently, 260 large area SDDs are employed to measure the trajectory of charged particles in the Inner Tracking System of the ALICE detector at the CERN Large Hadron Collider (LHC), as reported for example in ref. \cite{Vacchi_et_al_1991}.

\begin{figure}[h] \centering

\includegraphics[width=10 cm]{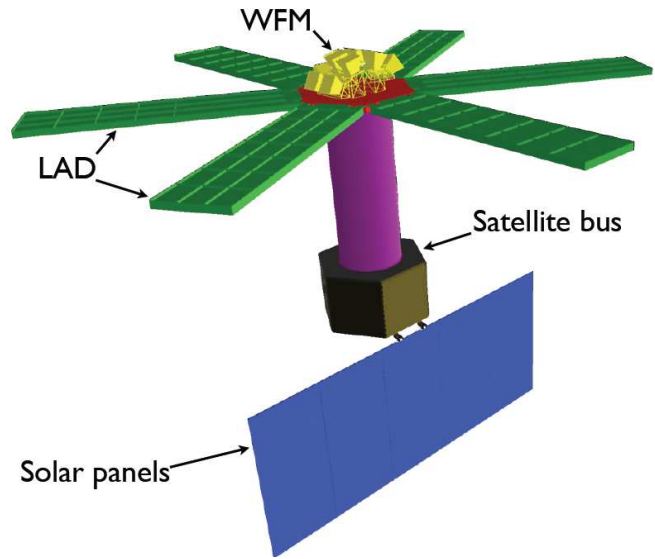}
\caption{Conceptual scheme of the LOFT satellite: Large Area Detector (green), Wide Field Monitor (yellow), optical bench (red), structural tower (purple), satellite bus (brown) and solar array (blue).}\label{fig:LOFT_design}

\end{figure}

SDDs are the sensors chosen for the two instruments aboard the {Large Observatory For x-ray Timing} (LOFT, see ref. \cite{LOFT_2012} and references therein): the Large Area Detector (LAD) and the Wide Field Monitor (WFM). The LAD is a collimated instrument for X-ray timing with the unprecedented geometric area of $\sim 15 \; \mathrm{m}^2$, corresponding to an effective area of $\sim 10 \; \mathrm{m}^2$ at an energy of 8 keV, a field of view of $\sim 1 \degree$ and a spectral resolution of $\sim 240$ eV Full-Width at Half Maximum (FWHM) at 6 keV (see ref. \cite{LAD_SPIE_2012} for more details). The effective area of the LAD is $\sim 15$ times larger than its predecessor, the Proportional Counter Array aboard the {Rossi X-ray Timing Explorer} satellite mission \cite{Jahoda_et_al_2006}. The WFM is an X-ray monitor based on the coded aperture imaging technique and is composed of five independent units, each one made of two co-aligned cameras, with an instantaneous field of view of $\sim 4.1$ sr. The WFM can simultaneously observe about one third of the sky with an angular resolution of $\sim 5$ arcmin and a location accuracy of $\sim 1$ arcmin for point sources (see ref. \cite{WFM_SPIE_2012} for further information). A sketch of the LOFT payload with the position of the main subsystems is shown in figure \ref{fig:LOFT_design}.

Silicon drift is one of the enabling technologies for LOFT because it allows the production of detector tiles with simultaneously a geometric surface larger than $\sim 70 \; \mathrm{cm^2}$ \cite{Rashevsky_et_al_2002}, {providing for the coverage of the LOFT geometric area with a relatively small number of units}, together with a spectral resolution of $\sim 200$ -- 250 eV FWHM for X-rays, and a collection time of $\sim 5 \; \mu$s \cite{Kushpil_et_al_2006}. The SDDs for LOFT are developed in collaboration with the Fondazione Bruno Kessler\footnote{\texttt{http://www.fbk.eu/}} (FBK) at Trento (Italy) under the XDXL and ReDSoX\footnote{\texttt{http://zoidberg.iaps.inaf.it/redsox/}} programs of research and development of the Italian Istituto Nazionale di Fisica Nucleare (INFN). The development programs started from the original design of the ALICE SDD sensors \cite{Rashevsky_et_al_2002} and aim at: (i) increasing the quantum efficiency for X-rays (at high energy by increasing the detector thickness and at low energy by reducing the thickness of the passive layers on top of the silicon bulk); (ii) modifying the passivation and metallisation layers on the surface to minimise the detector sensitivity to the environmental conditions; (iii) reducing the leakage current in order to improve the spectral resolution.

LOFT is one of the five satellite-borne missions selected by the European Space Agency\footnote{\texttt{http://sci.esa.int/science-e/www/object/index.cfm?fobjectid=48467}} (ESA) in 2011 for a three-year Assessment Phase study within the long-term science program ``Cosmic Vision 2015 -- 2025''. As detailed in the Yellow Book\footnote{\texttt{http://sci.esa.int/loft/53447-loft-yellow-book/}}, the key scientific objectives of LOFT can be grouped in three main topics: determine the equation of state of cold ultra-dense matter in neutron stars, measure the effects of strong field gravity down to a few gravitational radii around black holes and neutron stars, and perform with the LAD spectral and timing observations of a wide variety of X-ray sources. The second objective in the list will be accomplished through the LAD observation of relativistically broadened and skewed profiles of spectral lines (e.g. iron $K_{\alpha}$) in the flux of X-ray binary systems and active galactic nuclei \cite {LOFT_2012}. The goal is the accurate measurement of the broad line parameters to derive the black hole spin. Similar observations require a spectral resolution of $\sim 250$ eV FWHM in order to measure the variation of the shape of the spectral lines and disentangle all different emission and absorption contributions that may be present in the iron $K$ band (3 -- 8 keV).

The other instrument aboard LOFT is the WFM, a coded aperture imager devoted to monitoring the highly variable X-ray sky accessible to the LAD, and eventually to trigger repointings of the satellite to observe transient objects, newly discovered targets and interesting states of known sources. The electronic noise of the detectors and the read-out electronics affects both the imaging capabilities and the spectral resolution of the WFM.

The spectral resolution of the LOFT instrumentation depends on the electronic noise of the system composed of the detector and Front-End Electronic (FEE). In particular, due to the FEE design of the LAD and WFM, the leakage current of the detector bulk represents the parallel component of the electronic noise (see ref. \cite{Spieler_2005} for a general discussion and ref. \cite{Zampa_et_al_2011} for the application to the SDDs). It is known from previous studies that the leakage current of semiconductors increases due to the radiation damage of charged and neutral particles in orbit \cite{Lindstrom_et_al_2001}, and this effect is expected to worsen the spectral resolution of the LOFT instrumentation. The selection of the SDDs for a satellite mission motivated us to study the variation of the detectors scientific performance when exposed to the space environment and to define proper mitigation strategies.

The LOFT mission has a baseline duration of 4.25 years, three months of which represent the Commissioning Phase. Most of the other mission parameters have been selected in order to reduce the radiation damage of the SDDs \cite{LOFT_2012}. For example, the baseline orbit selected during the industrial study is a circular and Equatorial low-Earth orbit with an altitude of 550 km and an inclination of $0\degree$. The relevant baseline parameters of the LOFT mission are collected in table \ref{tab:LOFT_parameters} and some of them are discussed in section \ref{sec:estimation}.

\begin{table}[h!]
\centering
\caption{Relevant baseline parameters of the LOFT mission.}\label{tab:LOFT_parameters}
\begin{tabular}{|l|l|}
  \hline
  Mission duration                             & 4.25 years \\
  Orbit altitude                               & 550 km \\
  Orbit inclination                            & $0\degree$ \\
  Expected proton fluence                      & $3.0 \times 10^6 \; \mathrm{cm^{-2}}$ \\
  Expected equivalent fluence (1 MeV neutrons) & $1.4 \times 10^7 \; \mathrm{cm^{-2}}$ \\
  Expected Total Ionising Dose                 & $\leq 2$ krad(Si) \\
                                               & $\leq 1.16$ krad($\mathrm{SiO_2}$)\\
  Required average spectral resolution (LAD)   & $\leq 240$ eV FWHM \\
  Required Equivalent Noise Charge (LAD)       & $\leq 17 \; \mathrm{e^-}$ rms \\
  Operating temperature of the SDDs (LAD)      & $\leq -10 \; \degree$C\\

  \hline
\end{tabular}
\end{table}

In this paper we report about the measurement of the effect of the displacement damage on the SDDs in the LOFT context. After describing the basic mechanisms of the radiation damage on the SDDs in section \ref{sec:NIEL}, we briefly give in section \ref{sec:estimation} the context of the radiation environment for the LOFT mission. In section \ref{sec:experimental_set-up} we describe the irradiation strategy and the experimental set-up at the accelerator, and in section \ref{sec:Results} we report the results of the measurements. We discuss our findings and draw our conclusions in section \ref{sec:discussion_conclusions}.

\section{Displacement damage in silicon detectors}
\label{sec:NIEL}

The interaction of charged and neutral particles in detectors and electronic devices aboard satellite instrumentation can produce different types of radiation damage. The displacement damage and the Total Ionising Dose (TID) increase the detector leakage current: the displacement damage affects the bulk component while the TID influences the surface current. Consequently, both effects are expected to worsen the spectral resolution of the detector.

At the LOFT equatorial Low-Earth orbit, we estimate that a small TID is expected, of the order of 1 -- 2 krad, while the displacement damage can significantly increase the leakage current of the SDDs, as will be clarified in this section. Due to the limited amount of irradiation experienced by the LOFT SDDs, there will be a negligible effect on the bulk doping concentration \cite{Lindstrom_et_al_2001} that is estimated to be of the order of few ppm. As a consequence, the detector full depletion voltage will not change appreciably. The main concern is related to the increase of the leakage current because it will degrade the energy resolution and could have an impact on the scientific performance of the SDD. Also the charge collection efficiency is affected by the radiation damage. However, as reported in a forthcoming article, it will have no significant impact on the detector performance. For these reasons, in the following we will address only the damage related leakage current degradation.

\subsection{Increment of leakage current}

The development of tracking experiments for high energy charged particles at the CERN LHC required the study of the modifications produced by a large fluence of such particles in the detector materials, especially silicon. A large number of articles have then been published on this topic and the interested reader can find a comprehensive review, e.g. by the RD48 (ROSE) collaboration \cite{Lindstrom_et_al_2001}.

Charged and neutral particles interacting in a semiconductor detector produce displacement damage in the lattice, thus increasing the bulk leakage current. The increment $\Delta I$ of the bulk leakage current is given by \cite{Segneri_et_al_2009}

\begin{equation}\label{eq:Delta_I}
    \Delta I = \alpha \; \Phi_{\mathrm{eq}} \; V		
\end{equation}

\noindent where $\Phi_{\mathrm{eq}}$ is the particle fluence incident on the device and rescaled to the equivalent effect of 1 MeV neutrons, $V$ is the detector volume in which the leakage current is measured and $\alpha$ is a coefficient of proportionality measured in dedicated tests, also known as the ``current related damage rate'' \cite{Moll_et_al_1999}. The increase  $\Delta I$ is independent of the type of silicon material that is used, high or low resistivity, n- or p-type, high or low concentration of carbon and oxygen impurities, as specified in ref. \cite{Moll_et_al_2002}.

To minimise the leakage current, the LOFT detectors will operate at low temperature, thus we make the conservative assumption that only a negligible amount of annealing takes place in orbit. For this reason in eq. (\ref{eq:Delta_I}) we use the value $\alpha  = 11.1 \times  10^{-17} \; \mathrm{A \; cm^{-1}}$, measured in ref. \cite{Segneri_et_al_2009} at a temperature of $-50 \; \degree$C. With this value the expected increase of leakage current is normalised at a measurement temperature of 20 $\degree$C. All the current measurements reported in this article follow this normalization.

\subsection{NIEL scaling hypothesis and hardness factors}

Apart from the proportionality to the fluence described in eq. (\ref{eq:Delta_I}), the calculation of the displacement damage is based on the hypothesis that the damage is linearly dependent on the Non-Ionizing Energy Loss (NIEL), regardless of the particle type (NIEL scaling hypothesis). Usually the damage produced by a particle $\xi$ is related to a reference value, that of 1 MeV neutrons ($D_n(1 \; MeV)$ = 95 MeV mb), by introducing a hardness factor $\kappa_{\xi}(E)$

\begin{equation}\label{eq:kappa_xi}
    \kappa_{\xi}(E) = \frac{D_{d, \xi}(E)}{D_{n}(1 \; MeV)}		
\end{equation}

\begin{figure}[h!] \centering

\includegraphics[angle=90, width=10 cm]{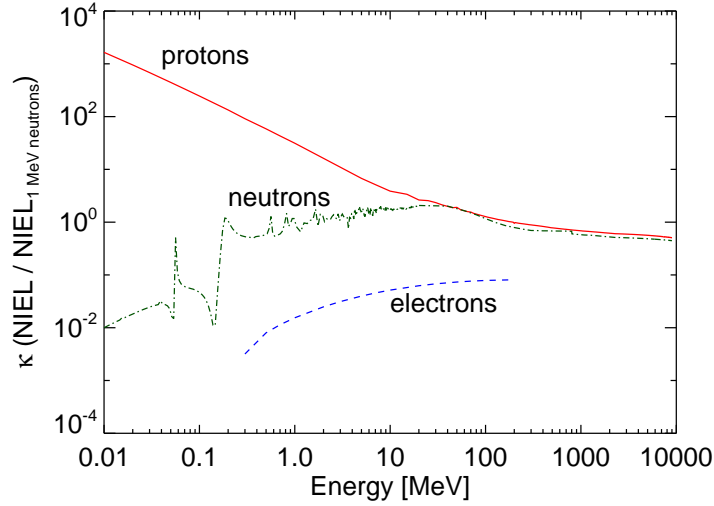}
\caption{Comparison of the $\kappa(E)$ factor for protons (red in color version), neutrons (green in color version) and electrons (blue in color version) at energies between $10^{-2}$ MeV and $10^4$ MeV, with data from ref. \cite{Vasilescu_Lindstroem_online}.}\label{fig:k_factor}

\end{figure}

\noindent The values of $\kappa(E)$ for various particles and energies are reported in literature (see e.g. ref. \cite{Spieler_2005, Vasilescu_Lindstroem_online}; see ref. \cite{Huhtinen_Aarnio_1993} for neutrons, protons and pions, ref. \cite{Konobeyev_1992} for neutrons, and ref. \cite{Summers_et_al_1993} for protons and pions). A plot with the comparison of the $\kappa(E)$ factor for protons, neutrons and electrons at typical energies of particles in orbit is shown in figure \ref{fig:k_factor}, with data from ref. \cite{Vasilescu_Lindstroem_online}\footnote{Online compilation in \texttt{ http://rd50.web.cern.ch/rd50/NIEL/default.html}}.

Combining eq. (\ref{eq:Delta_I}) and (\ref{eq:kappa_xi}), for the $\xi$-type particles the equivalent fluence $\Phi_{\mathrm{eq}}$ in eq. (\ref{eq:Delta_I}) is given by the product of the particle fluence $\Phi_{\xi}$ times the hardness factor $\kappa_{\xi}(E)$. Consequently, the current increment is given by

\begin{equation}\label{eq:Delta_I_kappa}
    \Delta I = \alpha \; \kappa_{\xi}(E) \; \Phi_{\xi} \; V		
\end{equation}

\noindent As shown in figure \ref{fig:k_factor}, for the energy range of the most abundant particles in Equatorial low-Earth orbit the NIEL from protons is highly dominant over neutrons and electrons. In fact, the integral flux of electrons, for example at an altitude of 600 km and an inclination of $5 \degree$, is comparable with protons only for energies around 1 MeV and rapidly decreases above $\sim 5$ MeV, but the $\kappa(E)$ factor of electrons in this energy range is more than three orders of magnitude lower than protons. One of the components of the radiation environment experienced by satellites in a low-Earth orbit is represented by the albedo neutrons produced in the interaction of the galactic and solar cosmic rays with the atmosphere. This phenomenon is influenced by the Earth magnetic field: a particle can reach a given point within the magnetosphere only if its momentum $cp$ normalized to its electric charge $q$ (called rigidity, $R = cp / q$ where $c$ is the speed of light) exceeds a threshold defined by the local intensity of the field (see for example ref. \cite{Herbst_et_al_2013} and references therein). Due to the high cut-off rigidity in the orbits selected for LOFT, $\sim 10$ GV, the integral flux of neutrons with energy between 1 and 10 MeV is similar to the proton flux but the neutron hardness factor is $\sim 30$ times smaller (at 1 MeV) to $\sim 3$ times smaller (at 10 MeV), as shown in figure \ref{fig:k_factor}. At higher energy the neutron flux rapidly decays, thus the contribution to the displacement damage is even smaller. For these reasons, we focus on the evaluation of the displacement damage from protons, neglecting the other types of particles.

\subsection{Damage annealing}
\label{sec:annealing}

The experimental data reported in this article were measured at a temperature for which the annealing is relevant. The leakage current component produced by the displacement damage is changed by the thermal redistribution of the damage centers in the crystal. The effect of this phenomenon can be accounted for by introducing a temporal dependence of the current related damage rate $\alpha$ in eq. (\ref{eq:Delta_I}) and eq. (\ref{eq:Delta_I_kappa}). A suitable model for our purpose is reported in ref. \cite{Moll_et_al_2002},

\begin{equation}\label{eq:annealing}
    \alpha(t) = \alpha_0 \; e^{-t / \tau_1} + \alpha_1 - \alpha_2 \ln (t/\tau_2)
\end{equation}

\noindent whose parameters depend on the annealing temperature. For example, at room temperature (21 $\degree$C) the values are $\alpha_0 = 1.23 \times 10^{-17} \; \mathrm{A \; cm^{-1}}$, $\alpha_1 = 7.07 \times 10^{-17} \; \mathrm{A \; cm^{-1}}$, $\alpha_2 = 3.29 \times 10^{-18} \; \mathrm{A \; cm^{-1}}$, $\tau_1 = 1.4 \times 10^4$ min, and $\tau_2 = 1$ min \cite{Moll_et_al_2002}. Similarly to the increase in leakage current, a universal annealing after the irradiation is observed, independent of the detector silicon material \cite{Moll_et_al_1999}.

\subsection{Temperature dependence of the leakage current}

The leakage current of a semiconductor is mainly due to the thermal generation of electron-hole pairs inside the depleted silicon bulk and has an exponential dependence on temperature \cite{Spieler_2005},

\begin{equation}\label{eq:leakage_vs_T_1}
    I(T) \propto T^2 \; e^{-E_g /2k_B T}
\end{equation}

\noindent where $T$ is the absolute temperature, $E_g$ is the band gap (1.12 eV for silicon) and $k_B$ is the Boltzmann constant. In the temperature ranges of interest for LOFT, eq. (\ref{eq:leakage_vs_T_1}) gives a reduction of the leakage current of a factor of 2 approximately every 7 $\degree$C decrease in temperature. The leakage current produced by the displacement damage has the same behaviour with temperature as the intrinsic leakage current of the non-irradiated devices, as shown for example in ref. \cite{Ohsugi_et_al_1988,Meidinger_1999}, because it is originated by the same process, i.e. the thermal generation.

\section{The LOFT context}
\label{sec:estimation}

\subsection{Characteristics of the LOFT SDDs}

The SDDs designed for LOFT are one-dimensional detectors subdivided into two electrically independent halves, each one with a set of collecting anodes at the edge \cite{Rashevsky_et_al_2002}. The charge cloud produced after the interaction of an X-ray is transported toward the anodes by means of a constant electric field, sustained by a progressively decreasing negative voltage applied to a series of cathodes on the top and bottom faces of the detector, down to the anodes at $\sim 0$ V. The operative principle of an SDD is described in detail in ref. \cite{Gatti_Rehak_2005,LAD_SPIE_2012}, where figures of the charge drift mechanism and the distribution of the drift potential are shown. In LOFT the detector thickness is 450 $\mu$m and the drift region has a length of 3.5 cm.

The LAD is a collimated instrument without imaging capabilities and the SDDs of this instrument have a geometric surface of 12.08 cm $\times$ 7.25 cm, a sensitive area of 10.85 cm $\times$ 7.00 cm, and an anode pitch of 970 $\mu$m. The sensor volume read-out by each anode is thus $1.53 \times 10^{-2} \; \mathrm{cm^3}$.

Although the WFM is a one-dimensional coded aperture imager, as explained in ref. \cite{Campana_et_al_2011} the width of the charge cloud, produced by the interaction of a photon, is used as a rough estimate of the distance between the photon impact point and the collecting anode along the drift direction, orthogonal to the main coding direction. For this reason, the SDDs for the WFM have a pitch of 145 $\mu$m, to optimise the position resolution and the imaging performance at low energy. The geometric surface of the SDDs of the WFM is 7.74 cm $\times$ 7.25 cm, the sensitive area is 6.51 cm $\times$ 7.00 cm and the sensor volume read-out by an anode amounts to $2.28 \times 10^{-3} \; \mathrm{cm^3}$.

The interested reader may find a detailed description of the SDDs for LOFT in ref. \cite{Rachevski_et_al_2014}.

\subsection{Electronic noise}
\label{sec:SDD_electronic_noise}

Discussing in detail all the noise contributions to the spectral resolution of the detection system is beyond the scope of this paper and the interested reader is referred, for example, to ref. \cite{Spieler_2005} for a general discussion and ref. \cite{Zampa_et_al_2011} for the specific application to the SDDs.

The size of the charge cloud produced by the interaction of a photon in an SDD increases during the drift process \cite{Zampa_et_al_2011,Campana_et_al_2011}. In particular, for the LAD anode pitch of 970 $\mu$m, on average $\sim 40$ \% of the detected photons are read out by one anode (single-anode events) and $\sim 60$ \% by two anodes (double-anode events). Situations in which the charge cloud is shared among three anodes are less than $\sim 1$ \%.

The scientific requirement for the LAD is that, at the end of the nominal mission duration (4.25 years) {and for photons of 6 keV energy,  $\mathrm{FWHM} \leq 200$ eV for the single-anode events and $ \leq 240$ eV on average ($\sim 40$ \% of single- and $\sim 60$ \% of double-anode events)}. This is translated into a requirement on the electronic noise, represented by the Equivalent Noise Charge (ENC): $\sigma_{ENC} \leq 17 \; \mathrm{e^-}$ root mean square (rms) for the single events. The corresponding requirement for the WFM is that, after 4.25 years in orbit, $\sigma_{ENC} \leq 13 \; \mathrm{e^-}$ rms. This value corresponds on average to $\sim 300$ eV FWHM for a photon of 6 keV.

\subsection{Radiation environment}

For satellites in Equatorial low-Earth orbit (i.e. below the Inner Van Allen Belt), the Earth magnetic field is an effective shield against Galactic and Solar protons with energy below $\sim 10$ GeV. The majority of the proton flux is thus represented by the particles ``trapped'' in the Earth magnetosphere. Trapped protons are concentrated in a region centered above South America (roughly with geographic latitude between $0 \degree$ and $50 \degree$ S and geographic longitude from $90 \degree$ W to $40 \degree$ E) and commonly referred to as the South Atlantic Anomaly (SAA), where the Inner Van Allen Belt comes closer to the Earth surface dipping down to an altitude of $\sim 200$ km.

An additional contribution is represented by a low-energy and highly directional population of protons near the Geomagnetic Equator and at an altitude between 500 km and 1000 km \cite{Petrov_et_al_2008}. Here we limit our study to the radiation damage from the trapped proton component, while the radiation damage from low-energy protons will be the topic of a forthcoming article.

In order to evaluate the increase in leakage current in orbit from eq. (\ref{eq:Delta_I_kappa}), we estimate the fluence of trapped protons with the publicly available and web-based SPENVIS software package\footnote{\texttt{http://www.spenvis.oma.be/}} \cite{SPENVIS_2002,SPENVIS_2009,SPENVIS_2010}. In the estimation we adopt the AP8 model of trapped proton fluxes, developed at the NSSDC of the NASA/GSFC, considered as a standard by ESA \cite{ECSS-E-ST-10-04C} and used for example to evaluate the radiation environment of NASA satellites in both low-Earth orbit (e.g. Swift \cite{Ambrosi_et_al_2002} and Hubble Space Telescope \cite{Jones_2000}) and high altitude orbit (e.g. Chandra \cite{O'Dell_2000}). From the literature, the AP8 model is known to be accurate within a factor of $\sim 3$, as reported for example in ref. \cite{Armstrong_Colborn_2000a,Armstrong_Colborn_2000b}.

SPENVIS includes the flux of trapped protons (AP8 model) and electrons (AE8 model) in both Solar minimum and maximum conditions. The proton flux during a Solar minimum (AP8-MIN) is generally higher than in a Solar maximum (AP8-MAX). For example, for an orbit at 600 km altitude and $5 \degree$ inclination, the integral flux of AP8-MIN at 1 MeV is $\sim 7$ times higher than AP8-MAX, and $\sim 6$ times higher at 10 MeV. For this reason, in our simulation we use AP8-MIN as a worst case.

In the orbits with the lowest expected fluence, AP8 is highly sensitive to small variations of the altitude and inclination. In addition, the flux of AP8-MIN is formally zero at 550 km and $0 \degree$. For these reasons, for the altitude of 550 km we adopt as a worst case the value of the fluence estimated at $3.5 \degree$ inclination. From a direct comparison, the flux measured by the Standard Radiation Environment Monitor aboard the Proba-1 satellite \cite{Proba-1_SREM} at 550 km and $0 \degree$ is $\sim 11$ times higher than the proton flux predicted by AP8-MIN for an orbit at 550 km and $2 \degree$. To be conservative we applied a further margin of a factor of 2 to take into account the possible uncertainties in the models, and we used a total margin of 20 times the fluence to calculate the increase of leakage current of LOFT.

Specifically for the calculation of the NIEL, the contribution of trapped protons may be evaluated in SPENVIS as the equivalent fluence at a single energy, selected by the user, transported through different values of the shielding thickness in aluminum (assumed as spherically symmetric). In our estimation we adopt for the protons the energy default value of 10 MeV, for which $\kappa(E)$ in eq. (\ref{eq:Delta_I_kappa}) is 4.8, specifically calculated in ref. \cite{Segneri_et_al_2009} for a silicon detector of 450 $\mu$m thickness, the same value of the LOFT SDDs. The shielding materials around the SDDs of the LAD are described in ref. \cite{Campana_et_al_2013} and are on average equivalent to 3.3 mm of aluminum.

In this article we restrict our analysis of the LOFT proton environment to the baseline orbit selected during the industrial studies, with an altitude of 550 km and an inclination of $0 \degree$, and the more ``unfavourable'' orbit at 600 km and $5 \degree$. For the two orbits and the assumptions above, the expected proton fluence for the LAD, including the margin of a factor of 20, is $3 \times 10^6 \; \mathrm{cm^{-2}}$ at 550 km altitude (equivalent to $1.4 \times 10^7 \; \mathrm{cm^{-2}}$ of 1 MeV neutrons) and $\leq 3.5 \degree$ inclination, and $1.8 \times 10^9 \; \mathrm{cm^{-2}}$ at 600 km and $5 \degree$ (i.e. $8.6 \times 10^9 \; \mathrm{cm^{-2}}$ of 1 MeV neutrons). In this region of orbital parameters there is a steep radial dependence to the proton fluence. For example, from 550 km altitude and $\leq 3.5 \degree$ inclination to 600 km altitude and $5 \degree$, the equivalent fluence of 10 MeV protons increases of 600 times. The data presented here are for a mission duration of 4.25 years, assuming a reference start on 1 Jan 2023.

We estimated with the SHIELDOSE-2 package of SPENVIS the TID for the LAD instrument, assuming as a worst case the orbit with 600 km altitude and $5 \degree$ inclination. Including the margin of a factor of 20 on the proton fluence from AP8-MIN, we obtain a TID $\leq 2$ krad(Si). Considering the relevant conversion factor (1 rad(Si) = 0.58 rad($\mathrm{SiO_2}$) from ref. \cite{Claeys_Simon_2002}), this translates into $\leq  1.16$ krad($\mathrm{SiO_2}$). The effects produced by the TID on the SDDs are an increase of both the trapped charge in the silicon oxide layer and the interface traps (see e.g. ref. \cite{Oldham_1999}). While the first effect is beneficial, since it entails an increase in the punch-through voltage between the drift cathodes, the second may determine an increase of leakage current due to surface generation.

\section{Experimental set-up and strategy}
\label{sec:experimental_set-up}

We measured the increase in SDDs leakage current produced by the displacement damage in a campaign at the accelerator of the Proton Irradiation Facility (PIF) in the Paul Scherrer Institute\footnote{\texttt{http://pif.web.psi.ch/}} (PSI) in Villigen (Switzerland). In the campaign we irradiated two SDDs produced during the LOFT Assessment Phase: one detector of the XDXL1 production and one of the XDXL2 batch. The XDXL1 SDD has been irradiated in order to study the variation of the Charge Collection Efficiency (CCE) produced by the displacement damage, which simultaneously increased the leakage current. Here we only report the measurement of the increase in the device bulk current, while the variation of the CCE will be the topic of a forthcoming article.

\subsection{Characteristics of the SDDs under test}
\label{sec:characteristics_SDDs}

The XDXL1 and XDXL2 prototype detectors have the same thickness (450 $\mu$m) and drift length (3.5 cm) of the SDDs for LOFT and a geometric area of 5.52 cm (in the anode direction) $\times$ 7.25 cm (in the drift direction). Both sensors have been produced with two different values for the anode pitch, to resemble the two LOFT instruments. The SDD side with the largest pitch is then called ``LAD half'', the other one ``WFM half''. For the XDXL1 sensor the pitch values are 835 $\mu$m (LAD half) and 294 $\mu$m (WFM half). For the XDXL2 detector the pitch is 967 $\mu$m for the LAD half and 147 $\mu$m for the WFM half, representative of the values for LOFT. A summary of the relevant features of the irradiated detectors is listed in table \ref{tab:SDD_characteristics}, as compared to the ones of the LAD and WFM. More details about the development of the different models of SDDs for LOFT are reported in ref. \cite{Rachevski_et_al_2014}.

\begin{table}[h!]
\centering
\caption{Characteristics of the SDDs under test: XDXL1 and XDXL2. The characteristics of the SDDs in the LAD and WFM are shown for comparison. All the SDDs in the table have a thickness of 450 $\mu$m and a drift length of 3.5 cm.}
\begin{tabular}{|l|c|c|c|c|c|}
  \hline
  Model             & Sensitive           & Geometric          & Anode       & Number    & Anode  \\
  name              & area [cm$^2$]       & area [cm$^2$]      & pitch       & of anodes & volume  \\
                    &                     &                    & [$\mu$m]    &           & [cm$^3$] \\
  \hline
  XDXL1             & $4.35 \times 7.02$  & $5.52 \times 7.25$ & 294         & 148              & $4.6 \times 10^{-3}$  \\
                    &                     &                    & 835         & 52               & $1.4 \times 10^{-2}$ \\
                    &                     &                    &             &                  & \\
  XDXL2             & $4.35 \times 7.02$  & $5.52 \times 7.25$ & 147         & 296              & $2.3 \times 10^{-3}$ \\
                    &                     &                    & 967         & 45               & $1.5 \times 10^{-2}$ \\
                    &                     &                    &             &                  & \\
  LOFT LAD & $10.85 \times 7.00$ & $12.08\times 7.25$ & 970   & 112              & $1.5 \times 10^{-2}$ \\
  LOFT WFM & $6.51 \times 7.00$  & $7.74 \times 7.25$ & 145   & 448              & $2.3 \times 10^{-3}$ \\
  \hline
\end{tabular}\label{tab:SDD_characteristics}
\end{table}

\subsection{Experimental set-up at the Proton Irradiation Facility}
\label{sec:PSI_PIF}

The PIF was constructed for testing of spacecraft components under a contract between ESA and PSI. We irradiated our detectors at the low energy site of the facility, which can provide proton beams with a maximum flux of $\sim 5 \times 10^8 \; \mathrm{cm^{-2} \; s^{-1}}$ and an energy ranging in quasi-continuous mode between $\sim 6$ and $\sim 63$ MeV using degrader foils. More information on the facility is reported in ref. \cite{Hajdas_et_al_1996}.

The proton spectrum was centered at 11.2 MeV, the value available at the facility nearest to the energy used in SPENVIS (10 MeV), and the FWHM was $\sim 6$ MeV.

\begin{figure}[h!]\centering
\includegraphics[width=12 cm]{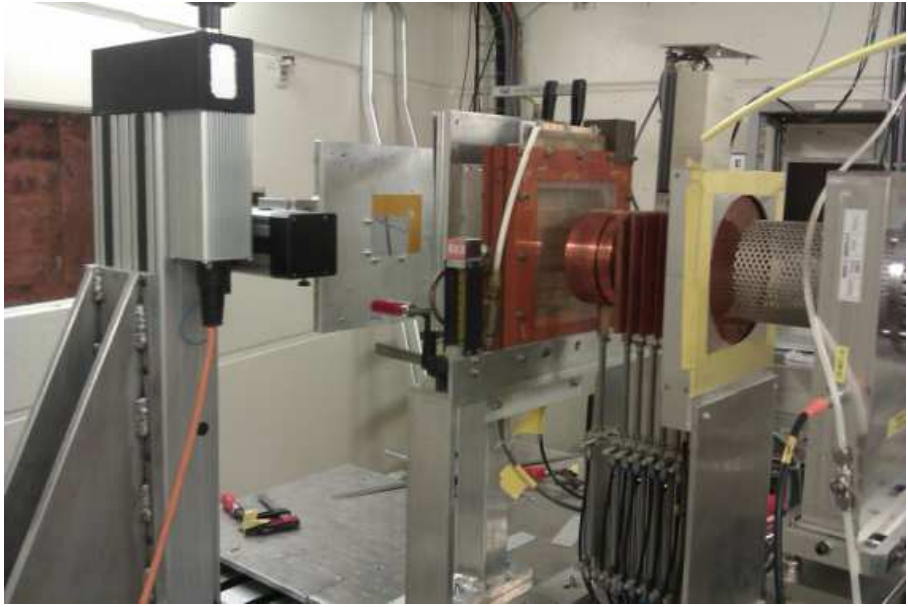}
\caption{Picture of the experimental set-up during the irradiation of the XDXL2 SDD. On the left the movable table and the sample holder are shown. The direction of the proton beam in the picture is from right to left.}\label{fig:PIF_PSI_set-up}
\end{figure}

The irradiation was performed in air and at room temperature (about 25 $\degree$C). A sample holder fixed on a movable table, which could be displaced in horizontal, vertical and longitudinal directions, was available at the facility. We specifically designed and produced sample holders to fix the device under test to the movable table. A laser cross-hair system was used as a reference to align the movable table with the geometrical center of the proton beam. A picture of the experimental set-up during the irradiation of the XDXL2 detector is shown in figure \ref{fig:PIF_PSI_set-up}.

Since we were interested only in the NIEL effects, the SDDs were not biased during the irradiation. The beam was kept normally incident to the detectors. The XDXL2 detector was irradiated without any shielding. The XDXL1 SDD was mounted on a Printed Circuit Board (PCB) with discrete components used as FEE and power distribution. In order to irradiate only the SDD and not the other electronic components, the PCB was shielded with an aluminum layer of $\sim 3$ mm thickness, able to stop all the protons in the beam, with a hole in the center to expose only the SDD.

\begin{figure}[h!]\centering
\includegraphics[angle = 90, width= 12 cm]{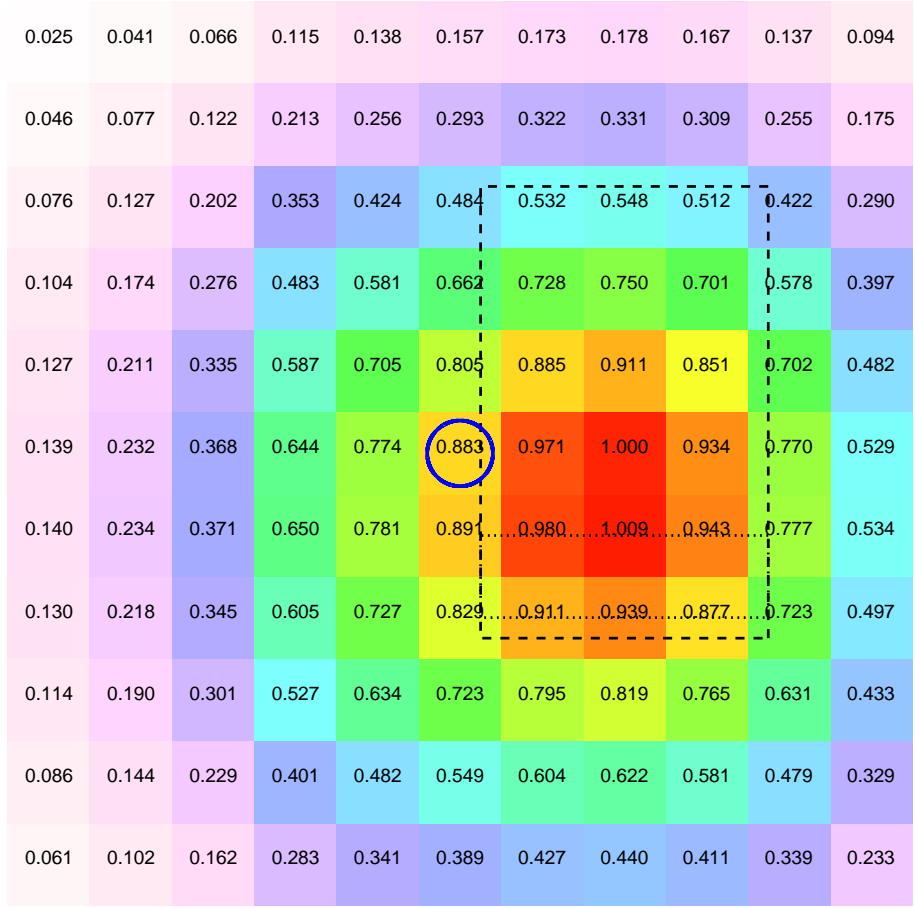}
\caption{Map of the reconstructed beam uniformity, in 1 cm steps, during the irradiation. The center of the beam is indicated by the blue circle. The horizontal $x$ axis (left to right) is along the direction of the charge drift. The vertical $y$ axis (up to down) is along the direction of the readout anodes. The dashed black box approximately defines the contour of the LAD half of the XDXL2 SDD, where the average fraction of the beam flux  is {81.4 \%.} The dotted black box approximately defines the contour (1 cm $\times$ 3.5 cm) of the XDXL1 anodes connected to the read-out electronics, where the average fraction of the beam flux is 88.8 \%.}\label{fig:beam_map}
\end{figure}

During the irradiation the beam was defocussed in order to cover the whole surface of the device under test. The beam profile had a gaussian shape. Before the irradiation we measured, with a motorised and remotely controlled scintillation detector, the proton flux with a step of 1 cm along both the horizontal ($x$ axis) and vertical ($y$ axis) directions, for a length of 11 cm on each axis. For each step the fraction of the flux with respect to the maximum was provided. We reconstructed the map of the beam fraction on the surface of 11 cm $\times$ 11 cm by multiplying the fractions along $x$ and $y$, as shown in figure \ref{fig:beam_map}. We found that the proton flux was displaced from the beam line center, being peaked at $x \simeq 2$ cm and $y \simeq 0$ cm.
The region around the maximum was also the region with the highest flux uniformity (see figure \ref{fig:beam_map}). For these reasons, during the irradiation we put the geometric center of the LAD half at $x \simeq 2$ cm and $y \simeq 0$ cm, in order to exploit the region with the maximum flux and the highest uniformity.

Two sets of wire chambers, one upstream and the other downstream with respect to the degrader foils, were used to monitor the beam flux during the irradiation. The counting rate of the two chambers was recorded every second. The time variability of the flux represented an error on the determination of the fluence. For this purpose, we assumed as fluence uncertainty the ratio between the standard deviation and the average value of the counting rate measured by the downstream wire chamber, the nearest one to the device under test. With this method, in each irradiation the estimated uncertainty was 2.4 \% of the total fluence.

\subsection{Fluence for the XDXL2 detector}

Since the detector leakage current was expected to be measured only about eight days after irradiation, when the residual damage due to the annealing at room temperature (21 $\degree$C) was predicted to be $\sim 41$ \% by eq. (\ref{eq:annealing}), we decided to expose it to a total fluence of $3.1 \times 10^9 \; \mathrm{cm^{-2}}$, corresponding to about 150 years in orbit at 600 km altitude and $5 \degree$ inclination. This fluence value does not include the margin of 20. In this way the current increase at the time of the first measurement was equivalent to an exposure of the LAD detectors to $\sim 60$ years in the radiation environment of such an orbit, allowing us to easily follow the subsequent damage annealing. In order to simplify the irradiation strategy, the whole fluence was provided in a single exposure. The flux during the irradiation was $3.7 \times 10^6 \; \mathrm{cm^{-2} \; s^{-1}}$.

The average fraction of the beam flux on the surface of the LAD half of the XDXL2 SDD was 81.4 \%, as highlighted by the dashed black box in figure \ref{fig:beam_map}, and the average fluence was $2.5 \times 10^9 \; \mathrm{cm^{-2}}$ (equivalent to $1.2 \times 10^{10} \; \mathrm{cm^{-2}}$ of 1 MeV neutrons). With this fluence, the silicon oxide of the SDD received a TID of {$\sim 1.5$ krad}($\mathrm{SiO_2}$), higher than the value expected for the SDDs in orbit (including margin).

\subsection{Fluence for the XDXL1 detector}

The XDXL1 SDD was irradiated to measure the variation of the CCE, whose annealing is slower than that of the leakage current \cite{Kramberger_et_al_2007}. For this reason we did not compensate for the annealing between the end of the irradiation and the measurements. With all these considerations, the fluence selected for the XDXL1 detector was $8.9 \times 10^8 \; \mathrm{cm^{-2}}$, equivalent to $4.3 \times 10^9 \; \mathrm{cm^{-2}}$ of 1 MeV neutrons and corresponding to ten times the value expected for 4.25 years in orbit at 600 km altitude and $5 \degree$ inclination (before the correction of the beam uniformity on the detector surface and without the margin of a factor of 20). The flux during the irradiation was half of the value used for the XDXL2 detector, $1.8 \times 10^6 \; \mathrm{cm^{-2} \; s^{-1}}$. Again we provided the whole fluence in a single exposure.

Eight contiguous anodes on the LAD half of the XDXL1 SDD are equipped with FEE based on discrete components. The contour of these anodes is highlighted by the black dotted box in figure \ref{fig:beam_map}. On these anodes the average fraction of the beam flux is 88.8 \% of the maximum.

\section{Irradiation results}
\label{sec:Results}

\subsection{Increment of leakage current after the irradiation}
\label{sec:measurement_XDXL2}

\begin{figure}[h!]\centering
\includegraphics[width = 8 cm, angle = 90]{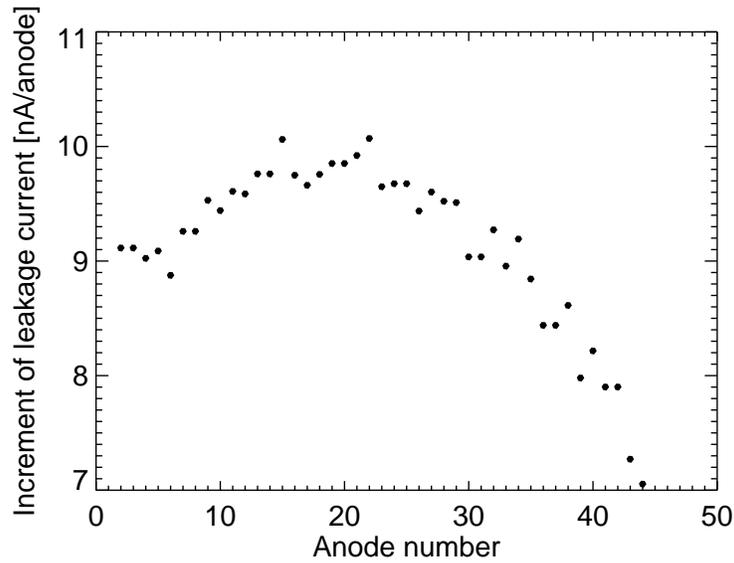}
\caption{First measurement of the increase in the anode leakage current on the XDXL2 SDD, performed at the probe station $\sim 6.3$ days after the irradiation and rescaled to a temperature of 20 $\degree$C.}\label{fig:leakage_current_XDXL2}
\end{figure}

The first measurement of the anode leakage current of the XDXL2 SDD was performed at the laboratory of INFN Trieste using a dedicated probe station $\sim 6.3$ days after the irradiation, following the method described in ref. \cite{Humanic_et_al_2003}. The average measured increase in the leakage current was 9.2 nA/anode at 20 $\degree$C, as shown in figure \ref{fig:leakage_current_XDXL2}. In the figure a clear profile of the measured increment values along the anodes is shown, maximum in the center and higher on the left side than on the right one. The trend mimics the variation of the beam fraction in the vertical direction in figure \ref{fig:beam_map}, which is maximum at the center of the LAD half and higher in the bottom part than in the top one.

Assuming the annealing at a constant temperature of 21 $\degree$C, $\sim$ 6.3 days after the irradiation the residual damage is 42.5 \% of the initial value. Applying eq. (\ref{eq:annealing}) to the irradiated fluence, and correcting for the non uniformity of the beam (on average 81.4 \% of the peak value), the expected increase of the leakage current at that time is 8.7 nA/anode. The difference between the measured and estimated value at the same time and temperature is thus 6 \%, and can be explained with the uncertainty on the estimation of the annealing from eq. (\ref{eq:annealing}) and the reconstruction of the average beam uniformity on the LAD half.

The value of the leakage current measured at the anodes after the irradiation includes the contribution of both the displacement damage and TID. Since we showed above that the measured increase is within 6 \% from the value estimated considering only the displacement damage, we conclude that the increment due to the TID is negligible, as expected.

\subsection{Damage annealing}
\label{sec:annealing_XDXL2}

\begin{figure}[h!]\centering
\includegraphics[width = 8 cm, angle = 90]{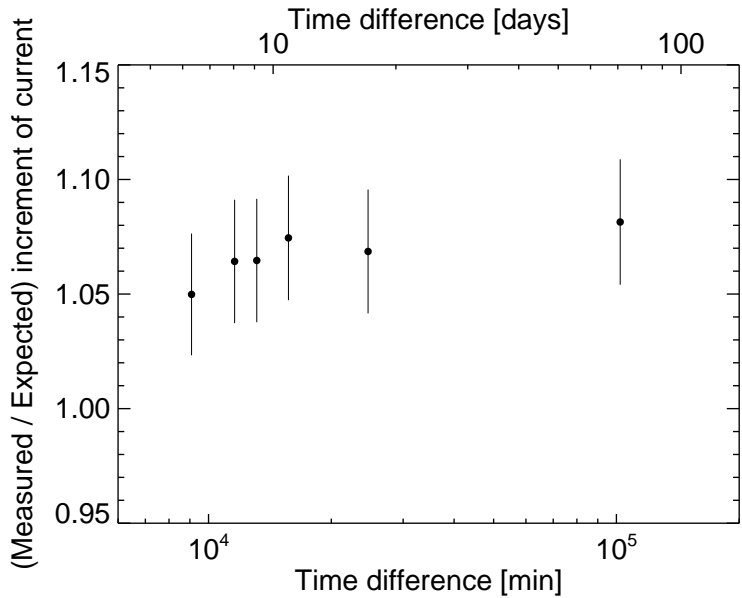}
\caption{Ratio between the measured and expected increment of leakage current as a function of time after the end of the irradiation.}\label{fig:annealing_XDXL2} \end{figure}

We repeated the measurement of the anode leakage current of the XDXL2 SDD with the probe station for a time interval between $\sim 6.3$ days and $\sim 70.9$ days after the irradiation, in order to study the damage annealing. In figure \ref{fig:annealing_XDXL2} we show the ratio between the measured current increment and the expected value as a function of time after the end of the irradiation. In the time interval of the measurements, the ratio between the measured and expected increment of current ranges between 1.05 and 1.08. The uncertainty on the ratio in figure \ref{fig:annealing_XDXL2} is dominated by the error on the fluence (2.4 \% estimated from the beam variability), since the error on the measured current (0.8 \%, based on the specification of the picoammeter \cite{Humanic_et_al_2003}) is negligible.

\subsection{Variation of the leakage current with temperature after the irradiation}
\label{sec:measurement_XDXL1}

Only a subset of eight anodes of the XDXL1 SDD is read-out by a FEE, based on discrete components. Each of these anodes is individually connected to a low gate capacitance SF-51 JFET (with capacitance of 0.4 pF) used as the input transistor of an Amptek A250F-NF charge sensitive amplifier. The feedback capacitor of 50 fF and a reset transistor are both integrated on the JFET die, leading to a reduction in the input stray capacitance and improvement in noise performance. A detailed description of the FEE circuit is reported in ref. \cite{Zampa_et_al_2011}.

The output ramps of all preamplifiers are compared with a high voltage threshold and the first comparator that fires (i.e. the one connected to the leakiest anode) starts the reset phase: all feedback capacitors are discharged to a lower voltage by activating a feedback loop built around the reset transistor. The period of the reset signal can be used to monitor the largest of the leakage currents of the eight connected anodes, following the formula \cite{Zampa_et_al_2011}

\begin{equation}\label{eq:reset_period}
    I = \frac{C_f \; \Delta V}{T_{reset}}
\end{equation}

\noindent where $I$ is the current (in pA/anode), $C_f$ is the feedback capacitance (50 fF), $\Delta V$ is the reset voltage (1.74 V), and $T_{reset}$ is the period of the reset signal (in ms).

\begin{figure}[h!]\centering
\includegraphics[angle = 90, width = 10 cm]{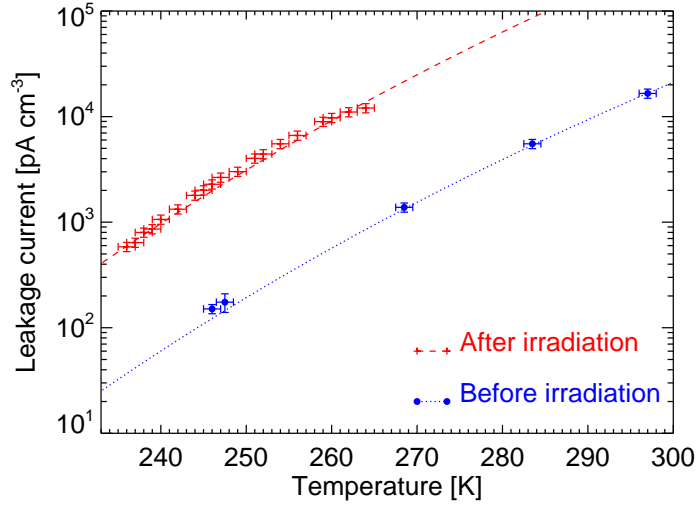}
\caption{Measurement of the leakage current of the XDXL1 detector after the irradiation (red curve) compared with the value before the irradiation (blue curve).}\label{fig:leakage_temperature_XDXL1}
\end{figure}

Before the irradiation, we measured the variation with temperature of the intrinsic leakage current of the XDXL1 SDD using the formula in  eq. (\ref{eq:reset_period}). The superposition of the current values before and after the irradiation is plotted in figure \ref{fig:leakage_temperature_XDXL1}. From the fit with the function in eq. (\ref{eq:leakage_vs_T_1}), in which the only free parameter is the normalisation $I_0$, we obtain $I_0 = (6.0 \pm 0.6) \times 10^8 \; \mathrm{pA \; K^{-2} \; cm^{-3}}$ (blue curve in figure \ref{fig:leakage_temperature_XDXL1}), corresponding to a leakage current of 0.2 nA/anode at 20 $\degree$C before the irradiation.

After the irradiation, we measured again the leakage current as a function of temperature with the same method as before. The fit with eq. (\ref{eq:leakage_vs_T_1}) yields $I_0 = (9.7 \pm 0.3) \times 10^9 \; \mathrm{pA \; K^{-2} \; cm^{-3}}$ (red curve in figure \ref{fig:leakage_temperature_XDXL1}) and the leakage current is 2.6 nA/anode. By subtracting the value before the irradiation we obtain a current increase of 2.4 nA/anode.

As shown in the map in Figure \ref{fig:beam_map}, the average fraction of the beam on the surface covered by the anodes connected to the FEE is 88.8 \% of the maximum. We measured the leakage current $\sim 5$ days after the end of the irradiation, when the residual damage was 44 \% of the original one. Applying eq. (\ref{eq:Delta_I_kappa}) using the current related damage rate given by eq. (\ref{eq:annealing}), the actual fluence and the $\kappa$ factor of 4.8 \cite{Segneri_et_al_2009}, the expected increment of the bulk leakage current is 2.4 nA/anode at 20 $\degree$C, in remarkable agreement with the measured value.

\section{Summary and conclusions}
\label{sec:discussion_conclusions}

At the PIF facility of the PSI accelerator in Villigen (Switzerland) we irradiated, with a proton beam centered at 11.2 MeV, an SDD of the XDXL1 batch and one of the XDXL2 production. The average intensity of the beam on the surface of the device under test was 88.8 \% for XDXL1 and 81.4 \% for XDXL2. The uncertainty on the proton fluence during the irradiation is 2.4 \%, resulting from the measured beam variability.

For both SDDs the proton fluence during the irradiation was more than an order of magnitude higher than the value estimated with SPENVIS for a mission of 4.25 years at the worst case scenario orbit of 600 km altitude and $5 \degree$ inclination, more severe than the baseline orbit for LOFT (550 km and $0 \degree$). For the provided fluence values, we estimated the increase of the SDDs leakage current basing on the NIEL approximation, using the damage parameter measured in ref. \cite{Segneri_et_al_2009} at $-50 \; \degree$C for a detector of the same thickness as our SDDs (450 $\mu$m), and the annealing model reported in ref. \cite{Moll_et_al_2002}.

Measuring the bulk leakage current following the method in ref. \cite{Humanic_et_al_2003}, we obtain on the XDXL2 detector an agreement within 6 \% of the measured and predicted increase. The value of the leakage current measured after the irradiation includes the contribution of both the displacement damage and TID. Considering the beam fraction on the SDD surface, during the irradiation the TID received by the oxide layer is {$\sim 1.5$ krad}($\mathrm{SiO_2}$), higher than the value expected for the LAD in orbit, including a margin of a factor of 20 on the fluence. Since the measured increase in leakage current is within 6 \% from the value estimated considering only the displacement damage, we conclude that the increment due to the TID is negligible. On the same detector we also find that, as expected, the annealing of the displacement damage after the irradiation follows the model in ref. \cite{Moll_et_al_2002}. We verified on the XDXL1 SDD that the leakage current after the irradiation follows the same trend in temperature as the intrinsic leakage current, before the irradiation. This is in agreement with the results published in ref. \cite{Ohsugi_et_al_1988,Meidinger_1999}.

We have shown in eq. (\ref{eq:leakage_vs_T_1}) how the leakage current is sensitive to the temperature. For this reason, the mitigation strategy against the displacement damage of LOFT in orbit is the decrease of the instrumentation temperature. Since the system weight and power required to decrease the LAD temperature with an active system (e.g. based on Peltier modules) would have been prohibitive due to the large detector surface, we instead selected a passive thermal control, based on the irradiation from dedicated aluminum radiators on the bottom face of the LAD panels. A similar system of radiators is employed on the WFM \cite{LOFT_2012}. Assuming in our calculations a mission duration of 4.25 years, the LAD scientific requirement of a spectral resolution better than 200 eV on single-anode events (corresponding to an electronic noise $\sigma_{ENC} \leq 17 \; \mathrm{e^-}$ rms) can be fulfilled at end of life with an operative temperature ranging from $-10 \; \degree$C for the orbit at 550 km altitude and inclination below $3.5 \degree$ down to $-52 \; \degree$C for an altitude of 600 km and an inclination of $5 \degree$. In this calculation we take into account both the soft (not reported here) and trapped proton components, the latter estimated using the worst case AP8-MIN model with a margin of a factor of 20. The industrial study during the LOFT Assessment Phase demonstrated that these operative temperatures can be reached with the passive thermal control system described above. In addition, at such temperature the annealing is negligible and cannot be exploited to reduce the leakage current.

\acknowledgments
LOFT is a project funded in Italy by ASI under contract ASI/INAF n. I/021/12/0, in Switzerland through dedicated PRODEX contracts and in Germany by the Bundesministerium f\"{u}r Wirtschaft und Technologie through the Deutsches Zentrum f\"{u}r Luft- und Raumfahrt under grant FKZ 50 OO 1110. The authors acknowledge the Italian INFN CSN5 for funding the Research and Development projects XDXL and REDSOX, and the INFN-FBK collaboration agreement MEMS2 under which the silicon drift detectors were produced. We gratefully acknowledge support of our measurements by the PIF team of PSI lead by W. Hajdas. The LOFT Consortium is grateful to the ESA Study Team for the professional and effective support to the assessment of the mission. This research has made use of NASA's Astrophysics Data System.

\end{document}